\documentclass{article}

\usepackage[english]{babel}

\usepackage[letterpaper,top=2cm,bottom=2cm,left=3cm,right=3cm,marginparwidth=1.75cm]{geometry}

\usepackage{amsmath}
\usepackage{graphicx}
\usepackage{graphbox}
\usepackage{bbm,amsfonts, amsthm,authblk}

\theoremstyle{theorem}
\newtheorem{theorem}{Theorem}
 
\theoremstyle{definition}
\newtheorem*{definition}{Definition}

\newcommand{\R}{\mathbb{R}}
\renewcommand{\S}{\mathbb{S}}
\newcommand{\Z}{\mathbb{Z}}

\newcommand{\inv}{^{-1}}

\newcommand{\simp}[1]{\{ #1 \}}
\newcommand{\ECC}{\mathrm{ECC}}
\newcommand{\ECT}{\mathrm{ECT}}
\newcommand{\SECC}{\mathrm{SECC}}
\newcommand{\SECT}{\mathrm{SECT}}

\usepackage{hyperref}
\usepackage{xcolor}
\hypersetup{
    colorlinks=true,
    linkcolor=violet,
    filecolor=violet,      
    urlcolor=violet,
    citecolor = violet
    }

\title{An Invitation to the Euler Characteristic Transform}
\author{Elizabeth Munch\footnote{Dept of CMSE and Dept of Mathematics, Michigan State University}}

\date{}

\begin{document}
\maketitle

\begin{abstract}

The Euler characteristic transform (ECT) is a simple to define yet powerful representation of shape. 
The idea is to encode an embedded shape using sub-level sets of a a function defined based on a given direction, and then returning the Euler characteristics of these sublevel sets. 
Because the ECT has been shown to be injective on the space of embedded simplicial complexes, it has been used for applications spanning a range of disciplines, including plant morphology and protein structural analysis.
In this survey article, we present a comprehensive overview of the Euler characteristic transform, highlighting the main idea on a simple leaf example, and surveying its its key concepts, theoretical foundations, and available applications. 
\end{abstract}

\section{Introduction}

Data can come in many forms and one of the aspects of that data that we might want to quantify is shape. 
In this case, and in many applications, we use the word ``shape'' to mean exactly what the non-mathematically trained layperson might think: How exactly does this object take up space in my 3D world? 
However, what might be simple to describe in a qualitative sense (smooth, bumpy, round, wiggly, long, etc) becomes particularly difficult to measure quantitatively. 
More intense still is finding ways to provide a value of similarity for the shapes being measured. 
Is this one more wiggly than the other? 
Are the smooth and bumpy ones more similar than the smooth and wiggly? 

Existing methods for encoding and comparing 3D shapes usually require additional information that may not be available. 
For instance, we may require that  the shape comes pre-annotated with landmarks (e.g.~\cite{Cates2017}). 
Alternatively, we start without this additional input but must find maps between the shapes to be able to relate different regions (e.g.~\cite{Boyer2011}). 
The former classes often require manual expert input which is both labor intensive and prone to bias; while the later is computationally intensive. 

In this paper, we focus on a newly available method of shape quantification and comparison built from the tools of topological data analysis (TDA) \cite{Dey2021,Amezquita2020,Munch2017}; that is, the use of tools from (Algebraic) topology to encode shape and structure in data. 
The construction we focus on in this paper is the Euler characteristic transform (ECT) \cite{Turner2014}, which encodes information about a shape in a way that does not require either maps between pairs of shapes to compare, or landmark information. 
And yet, it is still a complete representation of a shape and thus has been used in applications ranging from 
plants \cite{Amezquita2021}
to proteins \cite{Tang2022}, 
to cells \cite{Marsh2022}, 
to bones \cite{Wang2021a}.

\begin{figure}
    \centering
    \includegraphics[width = .7\textwidth]{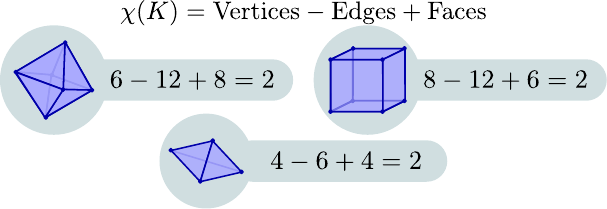}
    \caption{A figure showing the Euler characteristic  for the tetrahedron, cube, and octahedron. Since they are all homotopic to a sphere, the Euler characteristic is 2 in every case.}
    \label{fig:EC_Polytopes}
\end{figure}

In this case, the tool used from topology is arguably one of the simplest and elegant topological invariants: the Euler characteristic. 
Defined first in an unpublisehd manuscript by Francesco Maurolico in 1537 \cite{Friedman2018} but rediscovered and made more general by Leonard Euler in 1758 \cite{Euler1758}, the Euler characteristic provides a single integer as a measurement for a shape.  
In its simplest form, consider a polyhedron $K$ with vertices $V$, edges $E$, and faces $F$ such as the examples of Fig.~\ref{fig:EC_Polytopes}. 
The Euler Characteristic is defined to be the alternating sum 
$$
\chi(K)   = |V| - |E|  + |F|.
$$
This can be generalized to any triangulable space, in particular to our case of interest, those of simplicial complexes, which we will discuss in more detail later. 

The power of the Euler characteristic comes from the fact that it is a \textit{topological invariant}; that is, any homotopic spaces will have the same Euler characteristic. 
This can be seen through its connection to homology, another topological invariant measuring properties of the shape of a space. 
While we will try to avoid going down a rabbit hole to define homology in this paper, the Euler characteristic is equivalent to the alternating sum of the ranks of the homology groups \cite[Thm.~2.44]{Hatcher2002}.

We wish to use the Euler characteristic in computational and applied settings to automatically differentiate shapes. 
Unfortunately, the Euler characteristic is a rather coarse invariant in the sense that many spaces could have the same value and so cannot be distinguished by this number alone. 
Thus if one simply represents their data for which we wish to measure shape as only its Euler characteristic, we will likely do a poor job of classification and representation in any form. 

For this reason, we combine the Euler characteristic with a tool that can, in a mathematically provable way, distinguish these different shapes in 3D. 
The directional transform is a way of using information about an embedded shape to define a filtration which is a main building block in the field of TDA.  
Fixing a direction in $\R^3$, represented as a vector on the sphere $\omega \in \S^2$, we can define a function on the space $K \subseteq \R^3$ which has levelsets perpendicular to $\omega$. 
Retaining the portion of $K$ with function value below a particular $a$, we can record how the Euler characteristic changes as the value $a$ changes. 
In this way, we can use the Euler charactersitic spread out over multiple threshold values; and further, encode this information over any choice of $\omega$.
The result is the \textit{Euler characteristic transform (ECT)} which we will explore in this paper.

\section{The ECT}

To simplify statements and exposition, in this paper we will assume our input data is an embedded simplicial complex living in $\R^d$ for $d=2$ or $d=3$.\footnote{The most general results in this area work instead with $O$-minimal structures \cite{Dries1998}, and of course can be extended for higher $d$, but the generality is often not needed for our applied settings.} 
A simplicial complex is a combinatorial object consisting of building blocks of increasing dimension: vertices, edges, triangles, tetrahedra, and higher dimensional analogues. 
Specifically, a (geometric) $k$-simplex is the convex hull of $k+1$ affinely independent points in $\R^d$, denoted $\sigma = \simp{ v_0,\cdots, v_{k}}$.
A simplex $\tau$ is a face of a simplex $\sigma$ if the vertex sets are nested, i.e.~$\tau \subseteq \sigma$: in this case we write $\tau \leq \sigma$. 
A (geometric) simplicial complex $K$ consists of a set of geometric simplices in $\R^d$ such that 
(i) every face of a simplex in $K$ is also in $K$; and 
(ii) if two simplices $\sigma$ and $\tau$ are in $K$, then their intersection is either empty or a face of both. 
For the purposes of this work, particularly in the face of applications, we will assume that all simplicial complexes contain finitely many simplices. 
Further, we assume that the geometric realization of all complexes are contained in some finite radius ball, $|K| \subseteq B(0,R) = \{ x \in \R^d \mid \|x\| \leq R\}$.

\begin{figure}
    \centering
    \includegraphics[width = \textwidth]{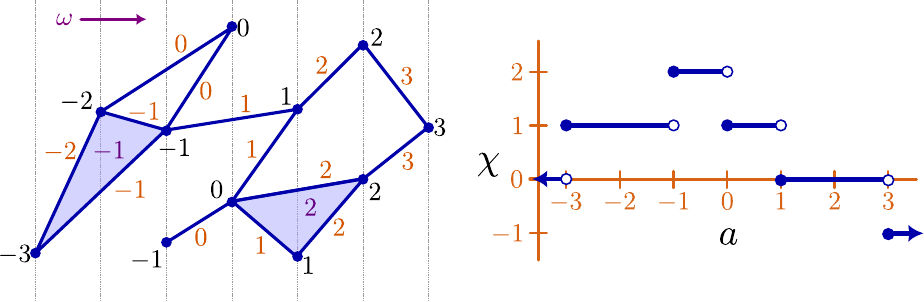}
    \caption{ At left we have a simplicial complex with simplex-wise function $f_\omega$ labeled for the horizontal choice of $\omega = 0$. 
    At right we have the corresponding Euler characteristic curve.  
    }
    \label{fig:DT_function}
\end{figure}

The main algebraic tool we will use here is that of the Euler characteristic. 
\begin{definition}
    The Euler characteristic of a simplicial complex $K$ is an alternating sum of counts of simplices in each dimension. Specifically, if $c_p$ is the number of $p$-simplices in $K$, then 
    \begin{equation*}
        \chi(K) = \sum_p (-1)^p c_p = c_0 -c_1+c_2-c_3 \cdots
    \end{equation*}
\end{definition}
Focusing on our case of simplicial complexes embedded in at most dimension 3, this simplifies down to 
\begin{equation*}
    \chi(K) = |\mathrm{vertices}| - |\mathrm{edges}| + |\mathrm{triangles}| - |\mathrm{tetrahedra}|. 
\end{equation*}
However, while homeomorphic spaces are promised to have the same Euler characteristic, this does not mean that non-homeomorphic spaces must have different Euler characteristics.
For example, the (empty) torus and the circle both have Euler characteristic 0 so cannot be differentiated by this number alone despite not being homeomorphic.
Thus, we combine the Euler characteristic with the \textit{directional transform}, defined next.
Given a simplicial complex $K$ in $\R^d$, we fix a direction $\omega \in \S^{d-1}$ thought of as a unit vector in $\R^d$. 
This choice of direction gives us a simplex-wise function 
\begin{equation*}
\begin{matrix}
   f_\omega : & K & \longrightarrow & \R \\
    & \sigma  & \longmapsto & \max_{v \in \sigma} \langle v, \omega \rangle 
\end{matrix}
\end{equation*}
where $\langle x, y \rangle = \sum x_i y_i$ is the standard dot product of the input vectors. 
See Fig.~\ref{fig:DT_function} for an example.
This function is set up so that we can study the \textit{sublevel sets} defined as
$$K_a = \{ \sigma \in K \mid f_\omega(\sigma) \leq a \}.$$ 
Note that our function $f_\omega$ is defined so that  $K_a$ satisfies the requirements of a simplicial complex for any $a$.

The \textit{Euler Characteristic Curve (ECC)}\footnote{The ECC for an arbitrary filtration function has a longer history than the ECT which is outside of the scope of this short paper; see e.g.~\cite{Heiss2017} and references therein. Note that the ECC is not actually a curve but a piecewise constant function. 
To the best of our knowledge, the first use of the term was in  \cite{Heiss2017} and arose from similar terms used in image processing and astronomy. Perhaps ``filtered Euler characteristic'' is a better term but at this point ECC is too deeply embedded in the literature for us to take a stand.} for direction $\omega$ is a function which returns the Euler characteristic for each sublevel set. 
Specifically, this is 
\begin{equation*}
\begin{matrix}
   \ECC_\omega: & \R & \longrightarrow & \Z \\
    & a & \longmapsto & \chi(K_a).
\end{matrix}
\end{equation*}
Again, see Fig.~\ref{fig:DT_function} for an example.
In practice, we have assumed that all simplicial complexes live in a ball of radius $R$; so for computational purposes we often restrict the domain to $ECC_\omega: [-R,R] \to \Z$.

\textit{A priori}, we do not know which direction to use to encode information about our particular simplicial complex $K$, so we put all of the directions together into a giant function called the Euler Characteristic Transform. 

\begin{definition}
Given a simplicial complex $K$ embedded in $\R^d$, the \textit{Euler Characteristic Transform (ECT)} is 
\begin{equation*}
\begin{matrix}
   \ECT(K): & \S^{d-1} & \longrightarrow & \Z^{\R} \\
    & \omega & \longmapsto & \ECC_\omega
\end{matrix}
\end{equation*}
where $\Z^{\R}$ denotes functions from $\R$ to $\Z$. 
\end{definition}

It has been shown that even though the Euler characteristic cannot distinguish between many non-homeomorphic spaces, the ECT with the added geometric information can.

\begin{theorem}[\cite{Curry2022,Ghrist2018}]
\label{thm:injectiveECT}
    The ECT is injective on the space of constructible sets in $\R^d$. 
    Specifically, if $ECT(K) \neq ECT(K')$, then $K \neq K'$. 
\end{theorem}
The above theorem was first proved for simplicial complexes embedded in dimension up to 3 by Turner \textit{et al.}~\cite{Turner2014}. 
In this work, it is a corollary of their main result in which they are focused on the so-called \textit{persistent homology transform} (PHT) where each direction of the sphere results in a persistence diagram rather than an Euler characteristic curve. 
As persistent homology is also outside the scope of this article, we direct the interested reader to \cite{Dey2021} for further specifics on this front.
Subsequently, using the machinery of Euler calculus \cite{Schapira1995}, this theorem was extended concurrently by \cite{Ghrist2018} and \cite{Curry2022}  to nice enough subsets in any ambient dimension.

\subsection{A fern leaf example}
\begin{figure}
    \centering
    \includegraphics[width = .4\textwidth]{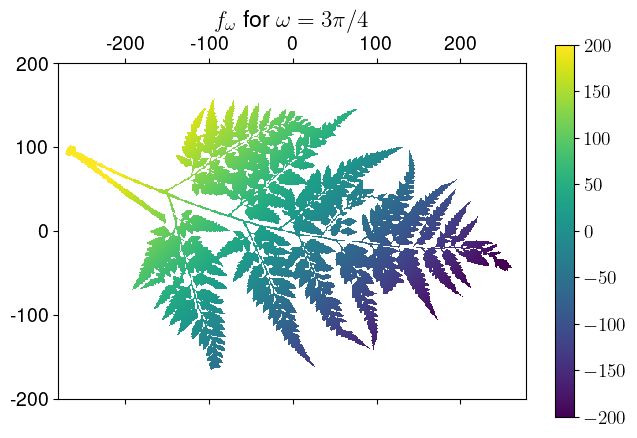}
    \includegraphics[width = .4\textwidth]{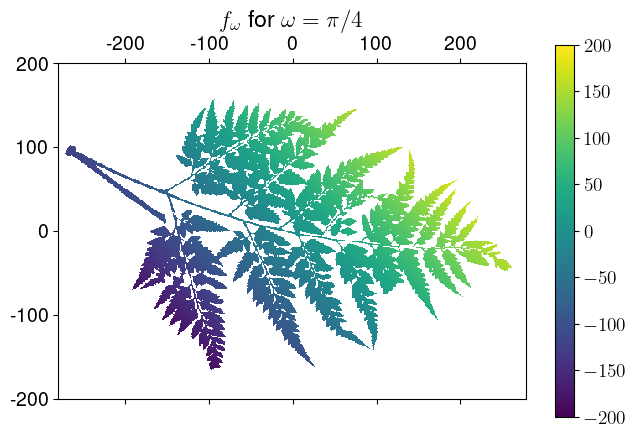}

\includegraphics[width = .15\textwidth]{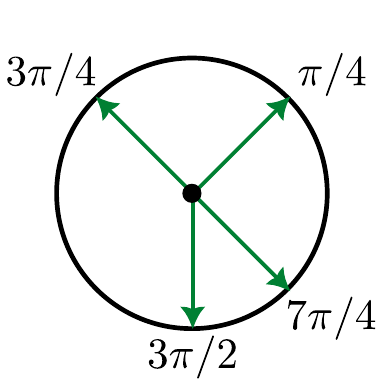}
    
    \includegraphics[width = .4\textwidth]{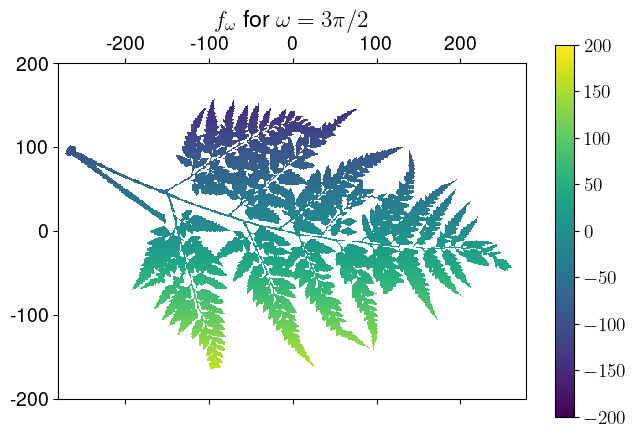}
    \includegraphics[width = .4\textwidth]{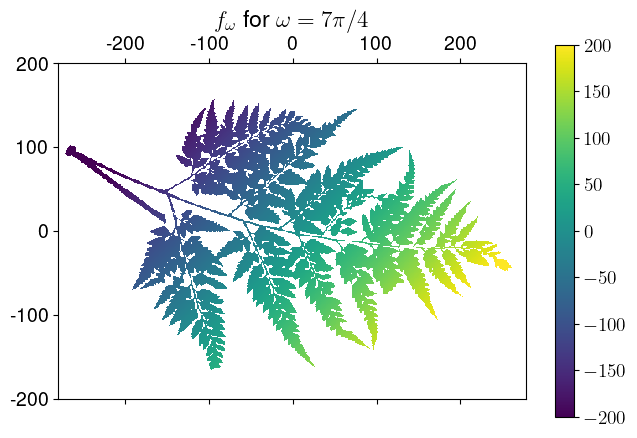}
    \caption{The function $f_\omega$ on the same fern leaf image is shown for $\omega = \pi/4$, $3\pi/4$, $3\pi/2$, and $7\pi/4$ where the function value of a pixel is given by color.}
    \label{fig:fernECT_Directions}
\end{figure}
We will show this procedure on the example of a binarized image of a fern leaf shown in Fig.~\ref{fig:fernECT_Directions} from \cite{White2020}. 
While technically speaking this input is binary pixel data (entries of 1 correspond to pixels included in the fern leaf), we can assume that we have triangulated each square included in the leaf and treat the subset of $\R^2$ as a simplicial complex.
We have also centered the image, and assume the coordinates of the pixels are relative to this origin.
In Fig.~\ref{fig:fernECT_Directions}, we see the induced function $f_\omega$ on this fern leaf for four different choices of $\omega$. 
Note that different directions induce different functions on the data where level sets are perpendicular to the direction chosen.

\begin{figure}
    \centering
    \includegraphics[width = .3\textwidth]{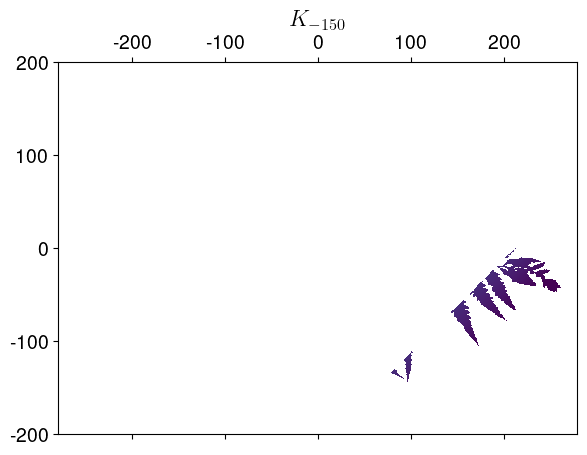}
    \includegraphics[width = .3\textwidth]{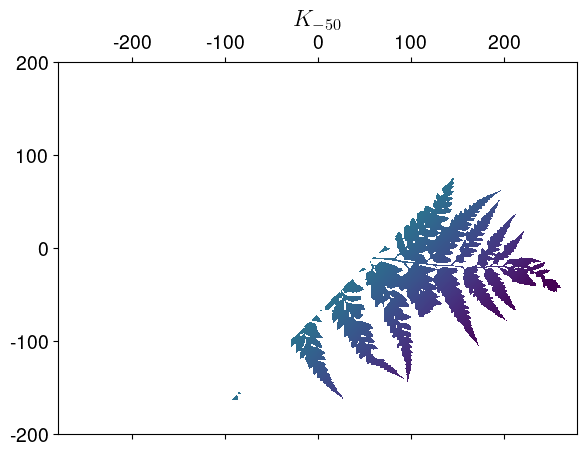}
    \includegraphics[width = .3\textwidth]{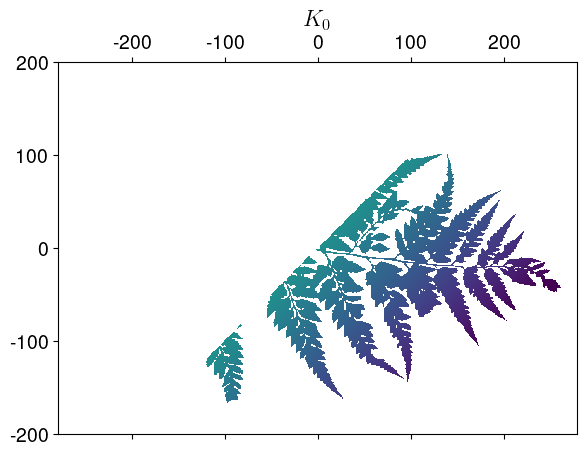}
    
    \includegraphics[width = .3\textwidth, align = c]{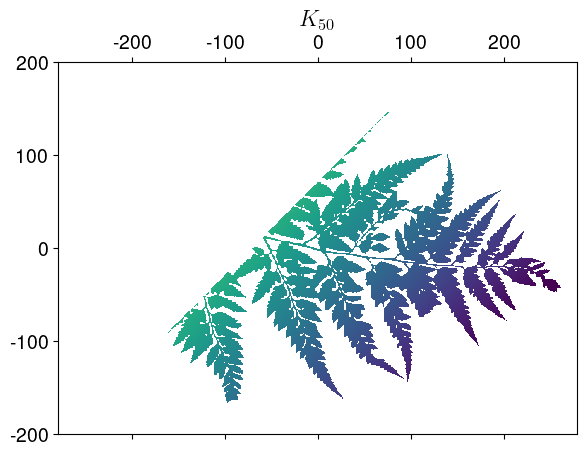}
    \includegraphics[width = .3\textwidth, align = c]{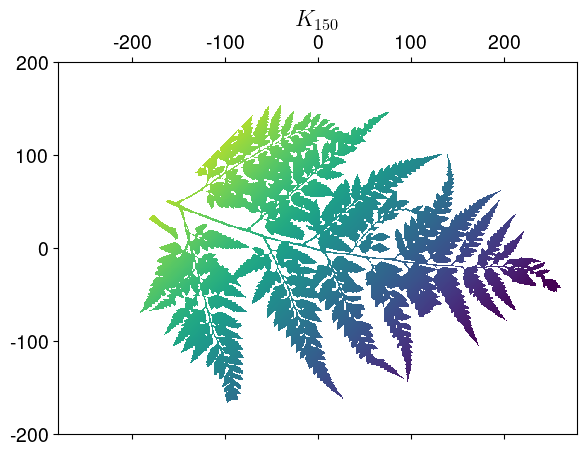}
    \includegraphics[width = .3\textwidth, align = c]{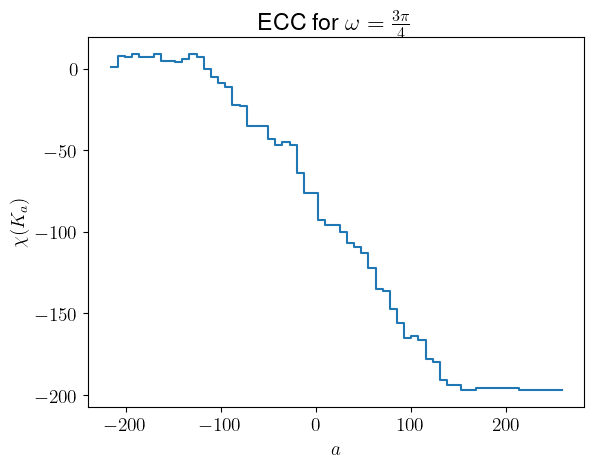}
    \caption{Fixing a direction $\omega = 3\pi/4$, the first five images are the portion of fern leaf with $f_\omega \leq a$ for $a \in \{ -150,-50,0,50, 150\}$. The full ECC for this $\omega$ is given in the last panel. }
\label{fig:fernECT_ECCs}
\end{figure}

Focusing for a moment on the case of $\omega = 3\pi/4$, we can look at the Euler characteristic of sublevel sets $\chi(K_a)$ for a few values of $a$. 
In Fig.~\ref{fig:fernECT_ECCs}, we have the sublevel sets for $a = -150,-50,0,50,150$. 
At each step in this filtration, we compute the Euler characteristic $\chi(K_a)$ and plot the value as a step function shown in the bottom left of the figure. 
The ECC starts close to zero, as we at first only see several connected components (positive small contribution from $\beta_0$) and few holes (negative small contribution from $\beta_1$). 
The function decreases until its final value of $\chi(K) = -197$ since the full fern leaf has only a few connected components (positive small contribution from $\beta_0$) but many holes (negative large contribution from $\beta_1$).
Note that the ECC for the fern in any direction will start at 0 when the function $a$ is lower than the minimum value of $f_\omega$, and stabilizes to $\chi(K)$ for the entire fern leaf when $a$ is above the maximum value of $f_\omega$. 

\begin{figure}
    \centering
    \includegraphics[align = c, width = .6\textwidth]{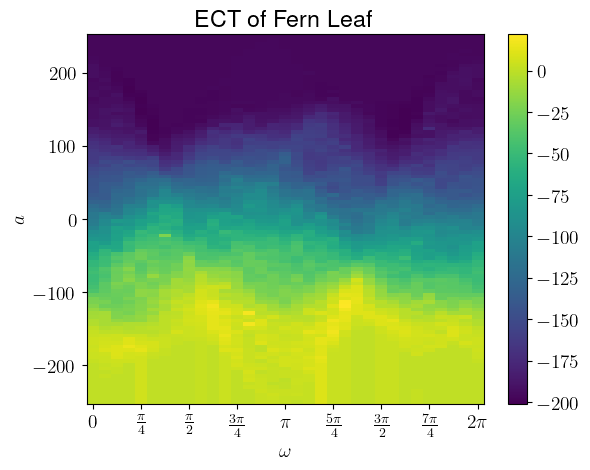}
    \includegraphics[align = c, width = .35\textwidth]{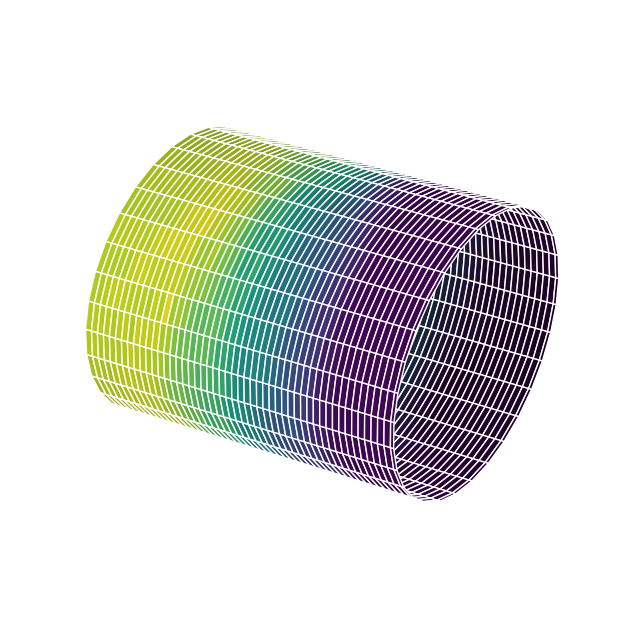}
    \caption{The ECT for the fern leaf is shown at left, with $x$-axis given by the direction $\omega$ in the circle; and $y$-axis given by the choice of threshold. The color gives the Euler characteristic for the thresholded fern $\chi(K_a)$ in the given direction. }
    \label{fig:fernECT}
\end{figure}

Since we can compute the ECC in any direction $\omega$, we can then compute the ECT by computing $\ECC_\omega$ for a discretized collection of directions. 
In the example of Fig.~\ref{fig:fernECT}, we have 64 evenly spaced directions on the circle; and compute $\chi(K_a)$ for values from $-200$ to $200$. 
For this particular example, we see that the values start around 0 at the bottom non-monotonically decreasing until it is $-197$ at the top thresholds $a$, which is the Euler characteristic of the full fern leaf. 
The values stay positive the longest in the directions $\omega$ that are approximately parallel to the fern leaflets, e.g.~$\omega =  \pi/2$ for the bottom pieces and $\omega \simeq 11\pi/8$ for the top, which can be seen in the yellow peaks in the ECT matrix.

\section{Some modifications and applications }

As with so many simple yet powerful ideas, there is an increasing literature of variations on the ECT which we survey here.

\subsection{Some notes on matrix representations}

Note that at least up to coordinates, this matrix representation provided in  the previous section is dependent on the embedding in the fern leaf. 
We will get the same matrix output if the leaf is rotated about the origin up to reindexing the $x$-axis. 
However this does mean that if we wish to compare this ECT to a similar fern leaf, we would likely want to ensure that the inputs are aligned in the ambient space beforehand, using a tool such as Procrustes \cite{Gower2004}. 

Another thing to note with this example is that in drawing the matrices, we have aligned the entries with the threshold value $a$ so that each entry corresponds to $\chi(f_\omega^{-1}(-\infty,a])$. 
However, depending on the application, a common modification is the following. 
Fix a number of thresholds to use, $T$. 
Then for a fixed direction $\omega$, the ECC is computed at $T$ equally spaced thresholds between the minimum and maximum function values on the shape for that fixed $\omega$. 
The matrix commonly used to represent the data in this setting has entry $M[i,\omega] = \chi(f_\omega^{-1}(-\infty,a_i^\omega])$ where $a_i^{\omega}$ is the $i$th threshold in direction $\omega$, but we do not necessarily have the same $a_i^{\omega}$ and $a_i^{\omega'}$ if $\omega \neq \omega'$. 
This later version is often useful in the case that we want to ignore global size differences; or in order to expand the amount of information stored in the matrix by not having long strings of constant entries in some directions. 

For example, in the application of Am\'ezquita \textit{et al.}~\cite{Amezquita2021}, the shapes to be studied were scans of barley seeds, all of which have a similar oblong shape with a characteristic indentation on one side. 
Prior to computation of the ECT, the seeds were aligned using PCA. 
The overall similarity of the shapes (as opposed to comparison of proverbial apples and oranges) meant that the ECT representation could be passed directly to ML algorithms as coordinate entries in the ECT aligned from one seed to the next.
Further, rather than encoding the ECT over equally spaced values over the entire bounding box, the representation used was equally spaced values over the maximum and minimum values in a fixed direction, leading to more compact representation of the shapes. 

\subsection{The SECT}

A variant of the ECT which is quickly gaining traction in applications due to its theoretical properties is the \textit{smooth Euler characteristic transform} (SECT), which was first defined in \cite{Crawford2019} and further investigated in \cite{Meng2022,Marsh2023}. 
The idea is to get away from the step-function nature of the ECT,  which causes issues with definitions of stability and access to statistical methodologies, by replacing the ECC with continuous functions. 

\begin{figure}
    \centering
    \includegraphics[width = .6\textwidth]{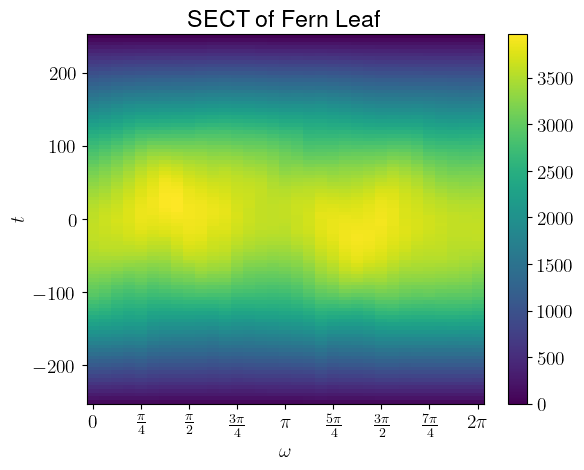}
    \caption{The SECT of the fern leaf example. In this case, each entry in the matrix at entry $(\omega,t)$ is $\int_{-R} ^t \left( \ECC_\omega(a) -  \overline{\ECC_\omega} \right) \, da$.  }
    \label{fig:Fern_SECT}
\end{figure}

First, we modify the ECT to be centered in the following sense.
Recall that all our simplicial complexes are assumed to be contained in a ball of radius $R$. 
Fixing a direction $\omega \in \S^{d-1}$, we can compute the average function value in this direction
\begin{equation*}
    \overline{\ECC_\omega} = \frac{1}{2R} \int_{-R} ^R \ECC_\omega(a) \, da.
\end{equation*}
We can replace the $\ECC_\omega$ with the 0-centered $a \mapsto \left ( \ECC_\omega(a) -  \overline{\ECC_\omega} \right)$. 
Fixing a direction, the \textit{Smooth Euler Characteristic Curve} (SECC) is defined to be 
\begin{equation*}
\begin{matrix}
   \SECC_\omega: & [-R,R] & \longrightarrow & \R \\
    & t & \longmapsto & \int_{-R} ^t \left( \ECC_\omega(a) -  \overline{\ECC_\omega} \right) \, da.
\end{matrix}
\end{equation*}
Using our assumptions, we can see that this function is 0 for the endpoints $t = \pm R$; or more generally for $t < -R$ and $t > R$ if we wish to extend the definition to $\R$.
Then the \textit{Smooth Euler Characteristic Transform} (SECT) is defined to be 
\begin{equation*}
\begin{matrix}
   \SECT(K): & \S^{d-1} & \longrightarrow & \R^{[-R,R]} \\
    & \omega & \longmapsto & \SECC_\omega.
\end{matrix}
\end{equation*}
The main reason for using this construction over the ECT is that the resulting functions have a Hilbert space structure, while still retaining injectivity results as an immediate consequence of Thm.~\ref{thm:injectiveECT}. 
The SECT of the Fern leaf from Fig.~\ref{fig:fernECT_Directions} discussed in the previous section can be seen in Fig.~\ref{fig:Fern_SECT}.

\subsection{Parameterized Families of ECTs} 

Another type of variation studied is that of the case of a collection of ECTs for an evolving family of shapes. 
In one example, the authors of \cite{Marsh2022} are studying time-lapse video of changing intestinal organioids. 
In their setting, there is a shape $K(s)$ for each time parameter $s \in [0,S]$. 
They define what they call the DETECT (DEtecting Temporal shape changes with the Euler Characteristic Transform) of the sequence as a function $[0,S] \to \R^{[-R,R]}$ given by
$$
s \mapsto \left( x \mapsto \int_ {\S^{d-1}} \delta_{x}\circ 
\SECC^{K(s)}_{\omega} \, d\omega \right). 
$$
Here, $\delta_{x}(f)= f(x)$ is the evaluation functional and $\SECC^{K(t)}_{\omega}$ is the smooth Euler characteristic curve for the shape $K(s)$ at time $s$. 
Specifically, this function gives the integral of the SECC over all directions but at a fixed height so that the resulting representation is orientation invariant. 

A related version of this problem is that studied in \cite{Kirveslahti2021}, where we start with a function $f:\R^d \to \R$.
Then they define the \textit{Lifted Euler Characteristic Transform} (LECT) by taking the ECT of a levelset of this function $f\inv(t)$:
\begin{equation*}
    \mathrm{LECT}(f)(v,h,t) 
    := \chi \left( 
    \{ x \in \R^d \mid x \cdot v \leq h, f(x) = t\}
    \right) . 
\end{equation*}
The superlevelset version, of course acronymed to be SELECT, is similarly defined as 
\begin{equation*}
    \mathrm{SELECT}(f)(v,h,t) 
    := \chi \left( 
    \{ x \in \R^d \mid x \cdot v \leq h, f(x) \geq t\}
    \right) . 
\end{equation*}
Then using the injectivity of Thm.~\ref{thm:injectiveECT}, all of these variants can be shown to be injective as well.

\subsection{Computational Considerations}
Computing the Euler characteristic transform is particularly appealing due to its computational complexity, as the needed information is an alternating sum of counts of simplices over a given direction. 
In this paper, we have computed examples using the \texttt{demeter}\footnote{\href{https://github.com/amezqui3/demeter}{github.com/amezqui3/demeter}} package for Python. 
An R-package called \texttt{SINATRA}\footnote{\href{https://github.com/lcrawlab/SINATRA}{github.com/lcrawlab/SINATRA}} is also available with a version specifically for protein data called \texttt{SINATRA-Pro}\footnote{\href{https://github.com/lcrawlab/SINATRA-Pro}{github.com/lcrawlab/SINATRA-Pro}}
Further computational speedups are available for streaming ECC data \cite{Heiss2017}, as well as using GPU capabilities \cite{Wang2022}.

Of course if this sort of construction is to be useful in applications, we need to be able to get away from the full mathematical generality of Thm.~\ref{thm:injectiveECT}. 
Specifically, for this construction to be an injective representation, we assume we have the full ECT for every direction $\omega \in \S^{d-1}$. 
But of course, this is not reasonable in practice since computers and image scans are restricted to finite representations of data.
In \cite{Curry2022}, they show a theoretical upper bound on the minimum number of directions $\omega_1, \cdots, \omega_m$ needed (\cite[Thm 7.14]{Curry2022}) to be estimated by 
\begin{equation*}
    m \geq (2b_\delta +1) (1+\tfrac{3}{\delta})^2 + O(\tfrac{3b_\delta}{\delta^2})^6
\end{equation*}
where $\delta$ controls the curvature, and $b_\delta$ is an upper bound for the number of critical values in the Euler characteristic curves in a $\delta$-ball of directions.
We will note that this bound is not tight, so there are many open questions about how to choose the number of directions in practice; and further if data driven sampling could be useful to ensure the directions used are focused on the ``interesting'' directions of the data.

\subsection{Reconstruction} 
With an injective representation such as the ECT, a common question, particularly from the domain scientists, is whether we can take a given ECT and reconstruct the original shape used to compute it. 
This is particularly useful when we want to be able to talk about things like an ``average'' shape. 
The lack of a well-posed inverse problem in other sorts of TDA constructions, such as the desire for a ``best'' homology class representative in persistent homology, often leads to difficulties with interpretation of the results for domain science questions. 
Because the ECT is injective, in theory this should be possible. 
Work to date has done this for planar graphs \cite{Fasy2019b}, simplicial complexes \cite{Fasy2019b}, greyscale images \cite{Betthauser2018},
and in related cases where we have augmented versions of the persistent homology transform \cite{Belton2020, Fasy2022a}. 
We also point the reader to the thesis of Sam Micka for an excellent introduction to the topic \cite{Micka2020}.

\subsection{Instances of ECT in deep learning }
Increasingly, recent work has begun to bring together the ECT and machine learning methods for classification and regression problems. 
For instance, the ECT representation of barley seeds in \cite{Amezquita2021} was passed to a support vector machine to classify the seeds by genetic line. 
A recent preprint by Paik \cite{Paik2023} opens the door to bringing the power of graph neural nets to classification via the ECT. 
Another recent preprint \cite{Nadimpalli2023} uses the ECT to build a loss function in the case of shape reconstruction problems.

\section{Conclusion} 

The Euler Characteristic Transform (ECT) is a powerful representation of shape. 
The basic tool, that of the Euler characteristic, is simple enough to explain to any mathematically curious person even while hiding deep mathematical underpinnings. 
We suspect that this construction could make for interesting high school or undergraduate student projects, particularly if the students have some basic coding abilities. 
However, even more so, it has the potential to be applied to many more shape analysis applications. 
In particular, its definition as an alternating sum of counts of simplicies makes for very fast computation, especially when held up against other computation times for tools in TDA.

The main drawback, which perhaps is the same as its injective strength, is that the ECT in standard formulations requires shape alignment prior to its use in learning pipelines. 
However, more work-arounds are being found, such as integrating out the dependence of the spherical coordinate in \cite{Marsh2022} or the use of rotation invariant neural networks in \cite{Paik2023}. 

One interesting future direction is that of using the directional transform with other sorts of representations of shape. 
Of course, the original definition \cite{Turner2014} was related to using the persistence diagram as the representation in each direction. 
However other sorts of shape descriptors in TDA can be used in conjunction with filtrations of this form such as the extended persistence diagram \cite{Turner2022} or merge trees \cite{Wang2023}. 
Further, we note that this paper has been focused on the directional transform version combined with Euler characteristics. 
However there are many more cases where encoding evolving Euler characteristics are useful in the more general setting of function data \cite{Dlotko2022,Beltramo2022,Roy2020}. 
With potential for use in so many applications, we hope that this survey will make the Euler characteristic transform  construction even more accessible to the shape analysis community.

\paragraph{Acknowledgment.}
The ECT was computed using the \texttt{demeter} package by Erik Am\'ezquita available at \url{https://github.com/amezqui3/demeter}. 
The author thanks Erik Am\'ezquita,  Sarah McGuire, Brittany Fasy, Hubert Wagner, and Justin Curry for helpful discussions.
The work of EM is funded in part by the National Science Foundation through CCF-1907591, CCF-2106578, and CCF-2142713. 
 
\bibliographystyle{plain} 
\bibliography{ECT} 

\end{document}